\documentstyle[epsf,aps]{revtex}
\begin{document}

\newcommand{\mbh}{M_{\rm bh}}
\newcommand{\rh}{R_{\rm h}}
\newcommand{\dc}{\delta_{\rm c}}
\newcommand{\mh}{M_{\rm h}}
\newcommand{\ea}{{\em et al.}\,\,}

\draft
\title{Dynamics of Primordial Black Hole Formation}
\author{J.C. Niemeyer}
\address{University of Chicago, Department of Astronomy and
Astrophysics, 5640 S. Ellis Avenue, Chicago, IL 60637, USA}
\author{K. Jedamzik}
\address{Max-Planck-Institut f\"ur Astrophysik,
Karl-Schwarzschild-Str. 1, 85740 Garching, Germany} 
\maketitle

\begin{abstract}
We present a numerical investigation of the gravitational collapse of
horizon-size density fluctuations to primordial black holes (PBHs) during the
radiation-dominated phase of the Early Universe. The collapse dynamics of
three different families of
initial perturbation shapes, imposed at the time of horizon crossing,
is computed. The perturbation threshold for black hole formation,
needed for estimations of the cosmological PBH mass function, is
found to be $\delta_{\rm c} \approx 0.7$ rather than the generally
employed $\delta_{\rm c} \approx 1/3$ if $\delta$ is defined as
$\Delta M/\mh$, the relative excess mass within the initial horizon
volume. In order to study the accretion onto the newly formed black
holes, we use a numerical scheme that allows us to follow the
evolution for long times after formation of the event horizon. In
general, small black holes (compared to the horizon mass at the onset
of the collapse) give rise to a
fluid bounce that effectively shuts off accretion onto the black hole,
while large ones do not. In both cases, the growth of the black hole
mass owing to accretion is insignificant. Furthermore, the scaling of
black hole mass with distance from the formation threshold, known to occur in
near-critical gravitational collapse, is demonstrated to apply to
primordial black hole formation. 
\end{abstract}

\pacs{PACS numbers: 04.70.Bw, 04.25.Dm, 97.60.Lf, 98.80.Cq}

\section{Introduction}
Primordial overdensities seeded, for instance, by inflation or
topological defects may collapse to primordial black holes (PBHs) 
during early radiation-dominated eras if they exceed a
critical threshold  \cite{PBHorigin1,PBHorigin2}. This particular PBH
formation process, which is
examined in this paper, occurs when an initially super-horizon size region of
order unity overdensity crosses
into the horizon and recollapses. Among the potentially observable
consequences of PBHs, should they be produced in cosmologically relevant
numbers, are thermal effects due to the Hawking evaporation of small
PBHs (manifested in the gamma ray background or as a class of very
short gamma ray bursts \cite{cliea97}) or purely gravitational effects
such as gravitational radiation of coalescing binary PBH systems
\cite{nakea97} or contribution of PBHs to the cosmic density
parameter. Upper bounds on
these signatures strongly constrain the spectral index of the fluctuation power
spectrum on small scales \cite{carrea94}.

Recently, the possibility that stellar mass PBHs constitute
halo dark matter has received attention in the
context of the MACHO/EROS microlensing detections \cite{macho}.  It has been
suggested that during the cosmological QCD phase transition, occurring at an
epoch where the mass enclosed within the particle horizon, $\rh \sim
t$, approximately
equals one solar mass, PBH formation may be facilitated due to
equation of state effects manifest in a reduction of the PBH
formation threshold \cite{jeda97}.

Every quantitative analysis of the PBH number and mass spectrum
requires knowledge of the threshold parameter, $\dc$, (for the specific
definition used here see below) separating perturbations that form
black holes from those that do not, and the resulting black hole mass,
$\mbh$, as a function of distance from the threshold. In a simplified
picture of the formation process, where hydrodynamical effects
are only accounted for in a very approximate way,
the universe is split into a
collapsing region described by a closed Friedmann--Robertson--Walker
(FRW) space--time and an outer, flat FRW universe. For a radiation
dominated universe, it can be shown that this ansatz yields $\dc
\approx 1/3$, where $\dc$ is evaluated at the time of horizon crossing
\cite{carr75}. On dimensional grounds, the natural scale
for $\mbh$  is the horizon mass, $\mh \sim \rh^3$, of the unperturbed FRW
solution at the epoch 
when fluctuations enter the horizon. However, these estimates for
$\dc$ and $\mbh$ are valid only within the limitations of the
employed model which cannot account for the detailed nonlinear
evolution of the collapsing density perturbations.

In order to determine $\dc$ and $\mbh$ for various initial conditions,
we performed one-dimensional, 
general relativistic simulations of the hydrodynamics of PBH
formation. We studied three families of perturbation shapes 
chosen to represent generic classes of initial data, reflecting the
lack of specific information about the distribution and classification of
primordial perturbation shapes. Our numerical technique, adopted
from a scheme developed by Baumgarte \ea \cite{baum95}, is sketched
in Section \ref{numerics}, followed by a description of the
general hydrodynamical evolution of the collapse and the results for
$\dc$ (Section \ref{hydro}) and a discussion of accretion after the
PBH formation (Section \ref{accret}). Defined as the excess mass within the
horizon sphere at the onset of the collapse, we find $\dc \approx 0.7$
for all three perturbation shapes. A numerical confirmation of the
previously suggested power--law scaling of $\mbh$ with
$\delta - \dc$ \cite{niejed97}, related to the well-known behavior of
collapsing space--times at the critical point of black hole formation
\cite{chop93}, is presented in Section \ref{scaling}. In this framework,
the PBH mass spectrum is determined by the dimensionless coefficient
$K$ and the scaling exponent $\gamma$, such that 
\begin{equation}
\label{scale}
\mbh = K \mh (\delta - \dc)^\gamma\,\,.
\end{equation}
We provide numerical results for $K$ and $\gamma$ for
the three perturbation families. These values may be used, in principle,
to determine PBH mass functions as outlined in \cite{niejed97}.

To introduce our numerical approach and isolate the dependence of
$\dc$ and $\mbh$ on the initial perturbation shape from the impact of
the equation of state, we restrict the discussion here to the purely
radiation-dominated phase of the Early Universe. In a separate
publication, we will investigate the change of $\dc$ before, during,
and after the cosmological QCD phase transition.  

Two other groups have, to our knowledge, published results of
numerical simulations of PBH formation in the radiation-dominated
universe \cite{Nade}. Our work differs
from theirs with regard to the numerical technique, the choice of
initial conditions, and the analysis of the numerical data. Wherever
possible and relevant, we compare our methods and results
with those previously published.

\section{Numerical technique}
\label{numerics}

The dynamics of collapsing density perturbations in the Early Universe
are fully described by the general relativistic
hydrodynamical equations for a perfect fluid, the field equations, the first
law of thermodynamics, and a suitable equation of state. 
We use a simple radiation dominated equation of state, $P=\epsilon
/3$, where $P$ is pressure and $\epsilon$ is energy density, 
as appropriate during most eras in the Early Universe.
The assumption of spherical symmetry is well justified for large
fluctuations in a Gaussian distribution \cite{barea86}, reducing the
problem to one spatial dimension. 

For our simulations, we have chosen the formulation of the hydrodynamical
equations by Hernandez and Misner \cite{hermis66} as implemented by
Baumgarte \ea \cite{baum95} (we omit restating the full system of equations
but instead refer to the equations published by Baumgarte \ea by a capital
``B'' followed by the respective equation number \cite{com1}). 
Based on the original equations by Misner and Sharp 
\cite{missha64}, Hernandez and Misner proposed to exchange the
Misner--Sharp time variable, $t$, with the outgoing null coordinate,
$u$. The line element then reads (Eq. B27)
\begin{equation}
ds^2 = -e^{2\Psi}du^2 - 2e^\Psi e^{\lambda/2} du \, dA + R^2
d\Omega^2\,\,,
\end{equation}
where $e^\Psi$ is the lapse function, $A$ is the comoving radial
coordinate, $R$ is circumferential radius, and $d\Omega$ is the
solid angle element (cf. Eq. B2).
After the transformation, the hydrodynamical equations retain the Lagrangian
character of the Misner--Sharp equations but avoid crossing into the
event horizon of a black hole once it has formed. Covering the entire
space--time outside while asymptotically approaching the event horizon,
the Hernandez--Misner equations are perfectly suited to
follow the evolution of a black hole for long times after its
initial formation without encountering coordinate singularities. This allowed
us, in principle, to study the
accretion onto newly formed PBHs for arbitrarily long times (in
contrast to earlier calculations \cite{Nade}) and therefore predict
final PBH masses.  

The Lagrangian form of the Hernandez--Misner equations allows the
convenient tracking of the expanding outer regions in a comoving numerical
reference frame. It also provides a simple prescription
for the outer boundary condition, as explained below. The extremely
low ratio of baryon number to energy density in the Early Universe
requires a re-interpretation of the comoving radial coordinate, $A$, in
Eq. (B1) and the comoving rest mass density, $\rho_0$, in (B3). We
re-define $\rho_0$ as the number density of a conserved tracer
particle with the purpose to define the comoving coordinate $A$ as the tracer
particle number enclosed within $R$. The variable $e$ is then
defined as the energy per tracer particle number density, such that
the energy density is $\epsilon = e\rho_0$. In the ultra-relativistic
limit $e \gg 1$, allowing us to replace  ``$1+e$'' with
``$e$'' in Eq.s B3, B4, B6, B14, and B38. This way, the Lagrangian
coordinate $A$ can be scaled to order unity together with all other
variables, which is desirable for reasons of numerical stability. 

Given the definition of the radial grid coordinate, a suitable
discretization of $A$ must be found. Numerical accuracy dictates to
deviate as little as possible from an equidistant grid partition lest
numerical instabilities occurring on superhorizon scales severely
constrain the grid size (see below). On the other hand, since $\Delta
R \sim R^{-2} \Delta A$ in a constant density medium, spatial
resolution is concentrated near the outer grid boundary and is worst
near the origin (where it is needed most) in case of equidistant $\Delta
A$. As a compromise between accuracy and resolution, we use an
exponentially growing cell size of the form
\begin{equation}
\label{grid}
\Delta A_{i} = (1 + \frac{{\cal C}}{N}) \Delta A_{i-1}\,\,,
\end{equation}
where ${\cal C}$ is a constant and $N$ is the total number
of grid points. Based on the standard convergence tests for numerical
resolution, we used $N = 500$ and ${\cal C} = 12$ for the
results reported below. 

The canonical boundary conditions (B18 and B40) are imposed at the
origin, while the outer boundary is defined to match the exact
solution of the Friedmann equations for a radiation-dominated flat
universe. Hence, the pressure follows the analytic solution 
\begin{equation}
\label{out_bound}
P = P_0 \left(\frac{\tau(A_{\rm N})}{\tau_0}\right)^{-2}\,\,,
\end{equation}
where $\tau(A_{\rm N})$ is the proper time of the outermost fluid element
(identified here with the FRW time coordinate, $t_{\rm FRW}$) and
$P_0$ and $\tau_0$ are the initial values for pressure and proper
time. The time metric (lapse) functions in (B1) and (B27) are fixed at 
\begin{equation}
e^\Phi = e^\Psi = 1\,\,
\end{equation}
at the outer boundary in order to synchronize the coordinate times $t$
and $u$ with the proper time of an observer comoving with the 
outermost fluid element, $\tau(A_{\rm N})$, and thereby with $t_{\rm
FRW}$ (note that B40
synchronizes $u$ with a stationary observer at infinity which is a
meaningless concept in an expanding space--time).

Owing to the presence of the curvature perturbation, the space--time
converges to the flat FRW solution only asymptotically. Therefore, the
accuracy of imposing Eq. (\ref{out_bound}) at the outer boundary is
presumed to grow with the size of the computational domain, 
removing the grid boundary farther from the density perturbation. In
particular, it is desirable to keep the boundary causally unconnected
from the perturbed region for as long as achievable. The
hydrodynamical evolution ensuing the collapse is highly dynamical for
$t \lesssim 100\,t_0$ (Section \ref{hydro}), where $t_0$ is the FRW
time at the beginning of the simulation, corresponding to the light
crossing time of a comoving distance of approximately $9 \rh$. Explicit
numerical experiments showed good agreement of the
numerical solution at the outer grid with the exact FRW solution for
all relevant perturbation parameters if the grid reached out
to $R_{\rm max} \gtrsim 9 \rh$. Extending the grid to these large radii
proved to be a non-trivial task for the Misner--Sharp part of the
numerical scheme, needed to initialize the Hernandez--Misner
computation, as will be outlined below. 

As the initial data is most naturally assigned on a spatial
hypersurface at constant Misner--Sharp (or, equivalently, FRW) time
$t$, one must transform the hydrodynamical and metric variables onto a
null hypersurface in order to initialize the Hernandez--Misner
equations. This can be done numerically in the way described by
Baumgarte \ea \cite{baum95}: first, the initial conditions are given on a
$t = $ const hypersurface and evolved using the Misner--Sharp
equations. Simultaneously, the path of a light ray is followed from
the origin to the grid boundary and the state variables on the path
are stored. After the light ray has crossed the grid, the Misner--Sharp
computation is terminated and the stored state values are used as
initial data for the Hernandez--Misner equations.  
As a consequence, the Misner--Sharp calculation needs to be carried
out until an initialization photon starting at the center reaches the
outer grid boundary. 
For the aforementioned reasons, it is preferable to use a
super-horizon size
computational domain. Since the light travelling time over such a large
grid is larger than the dynamical time for collapse to a PBH ($\sim t_0$), a
black hole already forms during the Misner--Sharp  
calculation before the initialization of the Hernandez--Misner grid is
completed. In order to avoid a breakdown of the Misner--Sharp
coordinate system inside the event horizon, the evolution of
fluid elements is stopped artificially if curvature becomes
large. More specifically, we found that the Misner--Sharp equations are
numerically well-behaved if the inner grid boundary is defined as the
innermost mass shell where the
Misner--Sharp lapse function fulfills $e^{\Phi} \gtrsim 0.2$.
The inner boundary conditions are henceforth chosen as the
frozen-in state variables on this shell. Despite this modification of
the evolution
equations the final results of the Hernandez--Misner calculation are
unaffected if the initialization photon is far away from the
collapsing region at the time of the boundary re-definition, i.e., if
the black hole forms at late times during the Misner--Sharp
calculation. In all cases reported here, this condition is satisfied. 

The second complication that arises in the Misner--Sharp coordinate
system is related to superhorizon scales. In a nearly flat expanding
space--time the square of the coordinate velocity, $U^2$ 
(where $U=e^{-\Phi}\partial R/\partial t$), and the ratio of
gravitational mass to radius, $2m/R$, grow with increasing
$R$. Both terms cancel identically in an exact FRW universe
for all $R$. On scales much larger than the horizon, however, they
become much greater than unity, and thus numerical noise can lead to
significant errors in the radial metric function, 
$\Gamma = (1+U^2-2m/R)^{1/2}$ (cf. B12). The
positive feedback of errors in  $\Gamma$, $\rho_0$ (B13), and $m$ (B6)
leads to a numerical instability of the Misner--Sharp equations on
superhorizon scales. It can be controlled by solving
for $\Gamma$, $\rho_0$, and $m$ at each time step by iteration, 
and by imposing a very restrictive allowed
density change of $\Delta \rho_0/\rho_0 \le 5 \times 10^{-4}$ per time step. 

None of the above mentioned problems exist in the Hernandez--Misner
formulation by virtue of the time slicing along null surfaces:
avoidance of the central singularity is guaranteed by the formation of
an event horizon, and the superhorizon
instability cannot occur because every grid point lies, by definition,
on the horizon. Therefore, after the assignment of initial
data is completed, the integration of the fluid equations in
Hernandez--Misner coordinates is numerically stable for arbitrarily
long times and on arbitrarily large spatial domains. In order to
achieve reasonable accuracy, however, the time
step size must be restricted to values much smaller than the
Courant-Friedrich-Levy (CFL) condition. The reason is most likely found in
the numerical integration of the lapse function, $e^\Psi$, which is only
first-order accurate \cite{baum95}. This problem is most important for
collapse with initial conditions 
close to the threshold for black hole formation, which leads
to the formation of very small black holes and gives rise 
to very strong space--time curvature. Decreasing the time step generally
causes a decrease of resulting black hole mass, $\mbh$.
At the numerical resolution chosen for this problem (see Eq.
\ref{grid}), the code is unable to follow the formation of black holes
smaller than $\mbh \approx 0.1$ in units of the initial horizon
mass. An adaptive mesh algorithm may be necessary to resolve this
problem. In agreement with studies of critical gravitational collapse
(\cite{evans94,koike95}), our experiments indicate that only the
coefficient $K$ in Eq. (\ref{scale}) is affected by the time step
variation, while the scaling exponent $\gamma$ appears very robust. 
Nevertheless, these problems disappear rapidly with distance to the
threshold such that convergence for larger PBH masses was attained.

In addition to the test-bed calculations reported by Baumgarte \ea
\cite{baum95}, we verified the accuracy of the code including the modified
outer boundary conditions by simulating the evolution of a flat
unperturbed universe. Using the time step restriction described above, the
numerical results for all hydrodynamical variables differ from the
analytic FRW solution by less than $10^{-3}$. 

\section{Hydrodynamical evolution of collapsing fluctuations}
\label{hydro}

We have studied the spherically symmetric evolution of three families
of curvature perturbations. Initial conditions are chosen to be
perturbations in the energy density, $\epsilon$, in
unperturbed Hubble flow specified
at horizon crossing. The first family of perturbations is described by
a Gaussian-shaped overdensity that asymptotically approaches the FRW
solution at large radii,
\begin{equation}
\label{gauss}
\epsilon(R) = \epsilon_0 \left[1 + A \exp{\left(-\frac{R^2}{2(\rh/2)^2}\right)}\right] \,\,. 
\end{equation}
Here, $R$ is circumferential radius,  $\rh = 2 t_0$ is the horizon 
length at the initial cosmological time $t_0$, and $\epsilon_0 = 3/(32 \pi
t_0^2)$, yielding $\mh = (4\pi /3) \epsilon_0 \rh^3=t_0$ 
for the initial horizon mass of the unperturbed space--time. In the absence
of perturbations, cirumferential radius corresponds to what is
commonly referred to as proper distance in cosmology, $r_p=ar_c$,
where $a$ is the scale factor of the universe and $r_c$ is the comoving cosmic
distance. 

\begin{figure}
\epsfxsize=80mm1
\epsfbox{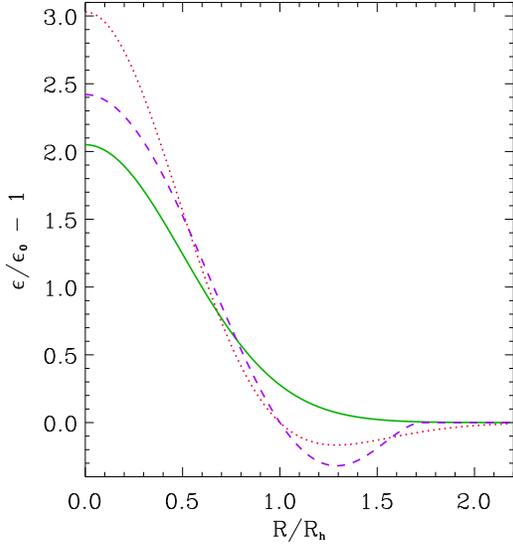}
\caption{\label{f0} Shapes of the critical perturbations as imposed at the
onset of the simulations: Gaussian (solid line, Eq. \ref{gauss}),
Mexican Hat (dotted line, Eq. \ref{mexhat}), and polynomial (dashed
line, Eq. \ref{poly}).}
\end{figure}

The other two families of initial conditions involve a spherical
Mexican-Hat function and a sixth order polynomial. These functions
are characterized by outer
rarefaction regions that exactly compensate for the additional mass of
the inner overdensities, so that the mass derived from the integrated
density profile is equal to that of an unperturbed FRW solution:
\begin{equation}
\label{mexhat}
\epsilon(R) = \epsilon_0 \left[1 + A \left(1 - \frac{R^2}{\rh^2}\right)
\exp{\left( -\frac{3 R^2}{2 \rh^2}\right)}\right] \,\,,
\end{equation}
and 
\begin{equation}
\label{poly}
\epsilon(R) = \left\{ \begin{array} {l@{\quad:\quad}l} 
\epsilon_0 \left[1 + \frac{A}{9} \left(1 - \frac{R^2}{\rh^2}\right)
\left(3 -\frac{R^2}{\rh^2}\right)^2\right]  & R < \sqrt{3} \rh \\
\epsilon_0 & R \ge \sqrt{3} \rh 
\end{array} \right.
\end{equation}
The amplitude $A$ is a free parameter used to tune the
initial conditions to sub- or supercriticality with respect to black
hole formation. The shapes of all three perturbations are illustrated in Figure
(\ref{f0}). 

The relevant dimensionless threshold parameter, $\dc$, for the purpose
of evaluating the cosmological abundance of PBHs is the energy overdensity in
uniform Hubble constant gauge averaged over a horizon volume
(i.e., synchronous gauge with uniform Hubble flow condition). It is
equivalent to the additional mass-energy  
inside $\rh$ in units of $\mh$. We find similar
values --- $\dc = 0.67$ (Mexican-Hat), $\dc = 0.70$ (Gaussian), and 
$\dc = 0.71$ (polynomial) --- for all three families of initial data in our
study, suggesting that the value $\dc \sim 0.7$ yields a more accurate
estimate for the cosmological PBH mass function than the commonly
employed $\dc \sim 1/3$. 

In cases where a black hole is formed, we define its mass, $\mbh$, as the
gravitational mass, $m$, enclosed by the innermost shell that conforms
to $e^\Psi \ge 10^{-10}$, the temporal evolution of all shells with
smaller $e^\Psi$ being essentially frozen in
(with regard to proper time of a distant observer). Owing to the steep
rise of $e^\Psi$ at the event horizon, the exact choice of the cutoff
value does not affect $\mbh$ within the accuracy reported in this
work. Unless otherwise specified, we henceforth quote $\mbh$ in units of
the initial horizon mass, $\mh$, and proper time in multiples of the initial
time, $t_0$. 

\begin{figure}
\begin{minipage}{80mm}
\epsfxsize=80mm
\epsfbox{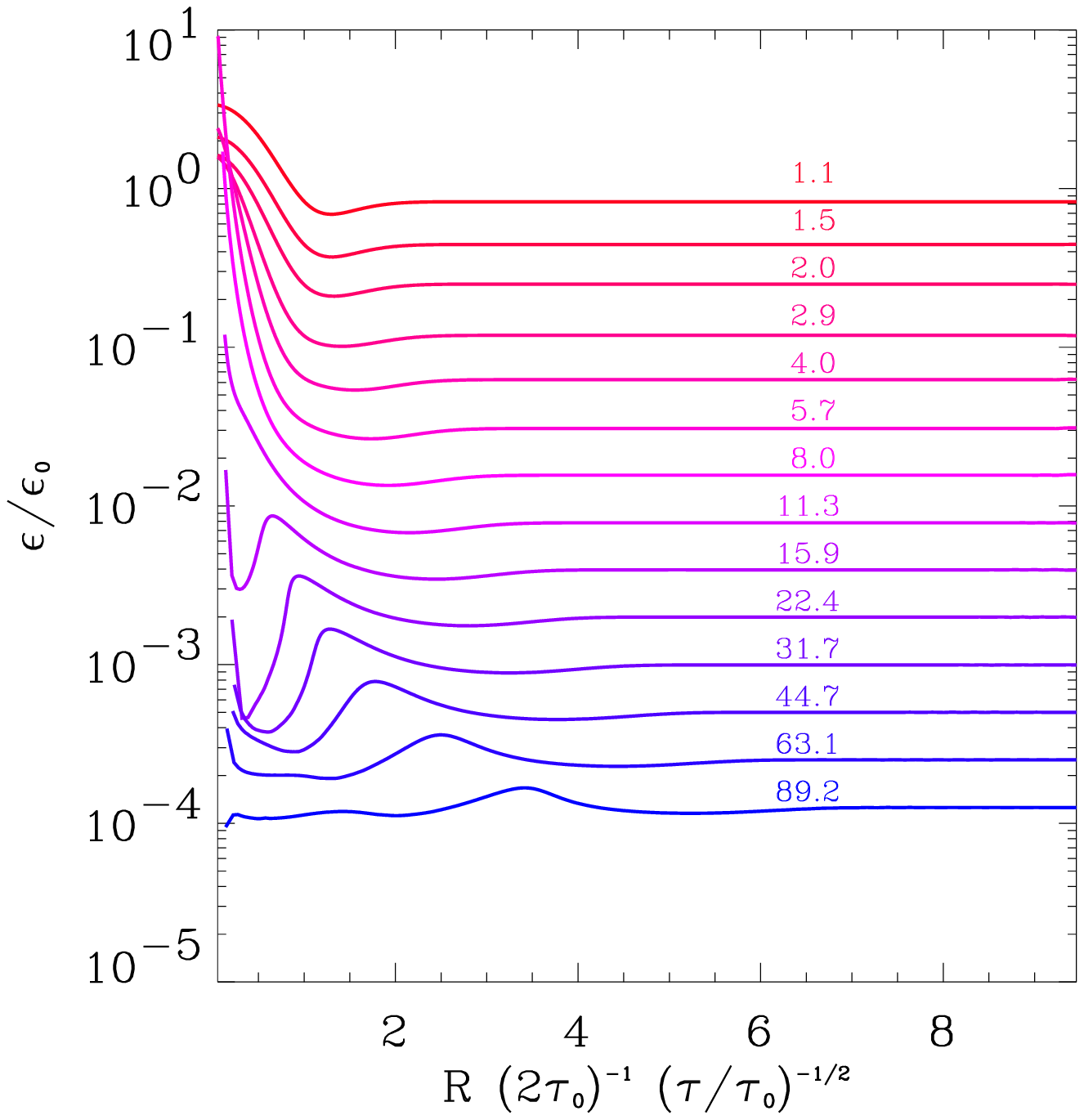}
\caption{\label{f1} Time evolution of a near-critical Mexican-Hat density
perturbation with initial $\delta = 0.6780$. A 
black hole with mass $\mbh = 0.37$ forms in the interior.}
\end{minipage}
\hspace{\fill}
\begin{minipage}{80mm}
\epsfxsize=80mm
\epsfbox{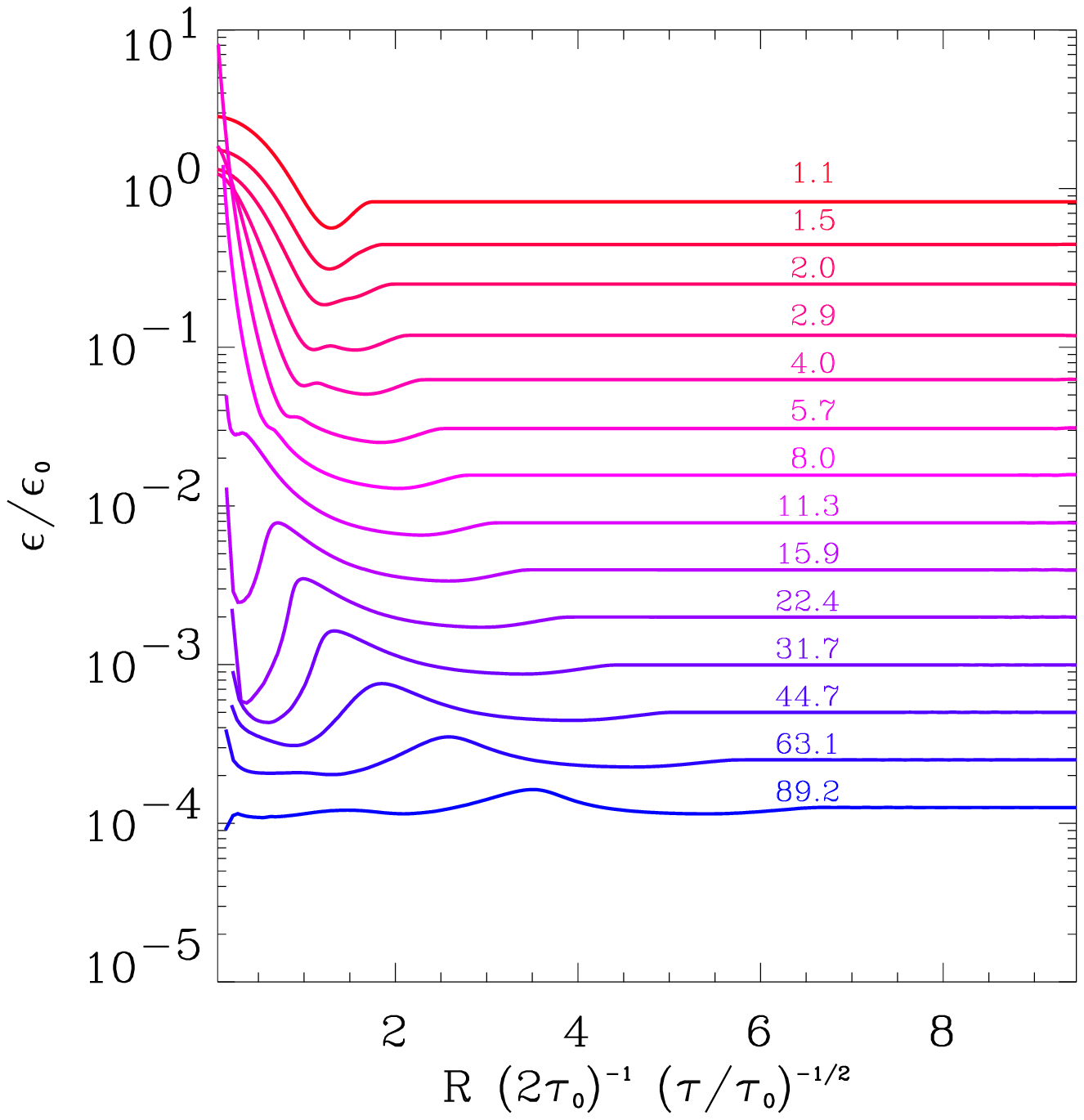}
\caption{\label{f2} Time evolution of a near-critical polynomial
perturbation with initial $\delta = 0.7175$, forming a 
black hole with mass $\mbh = 0.36$.}
\end{minipage}
\end{figure}
\begin{figure}
\begin{minipage}{80mm}
\epsfxsize=80mm
\epsfbox{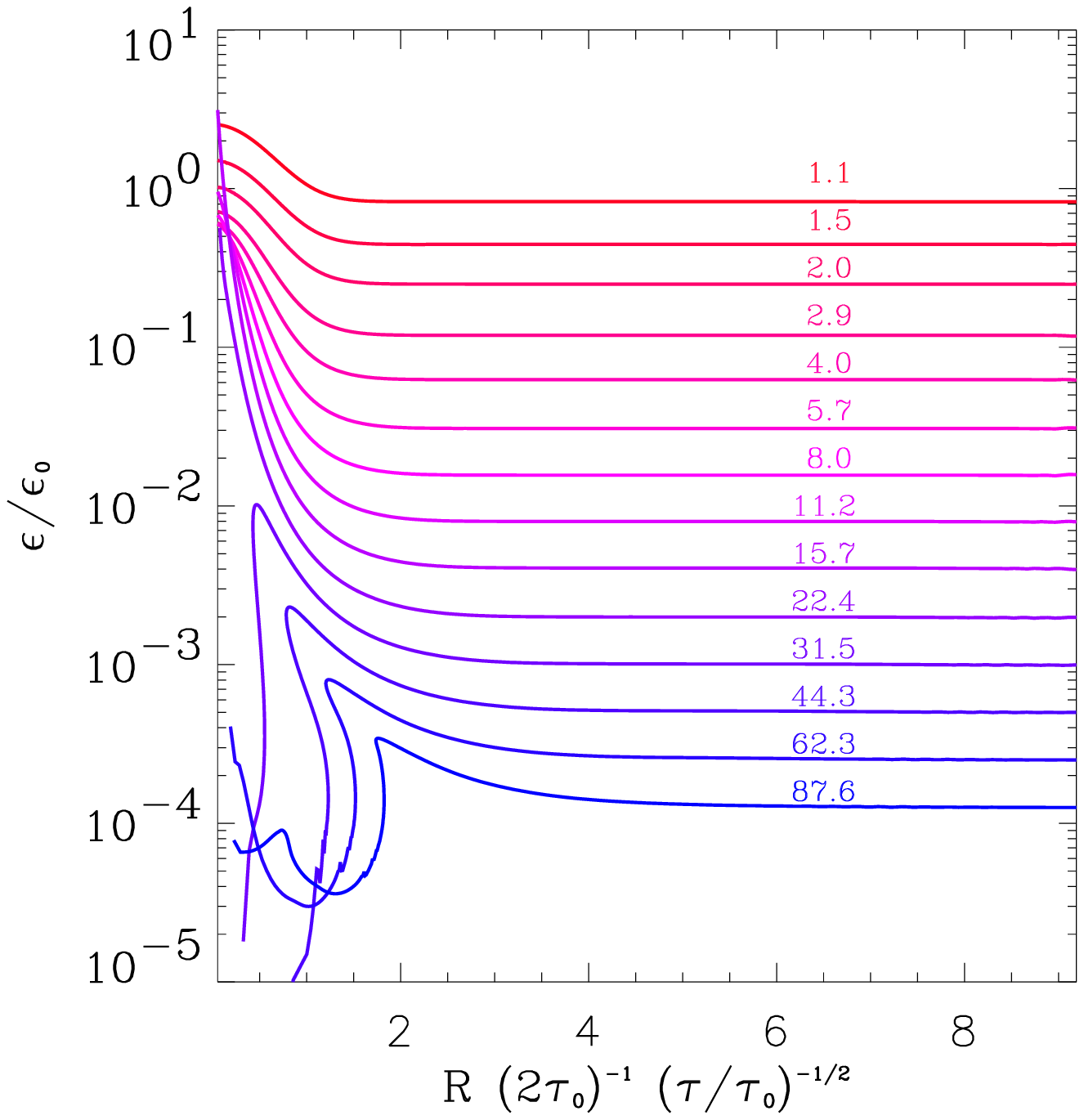}
\caption{\label{f3} Time evolution of a near-critical Gaussian-curve
perturbation with initial $\delta = 0.7015$, forming a 
black hole with mass $\mbh = 0.37$.}
\end{minipage}
\hspace{\fill}
\begin{minipage}{80mm}
\epsfxsize=80mm
\epsfbox{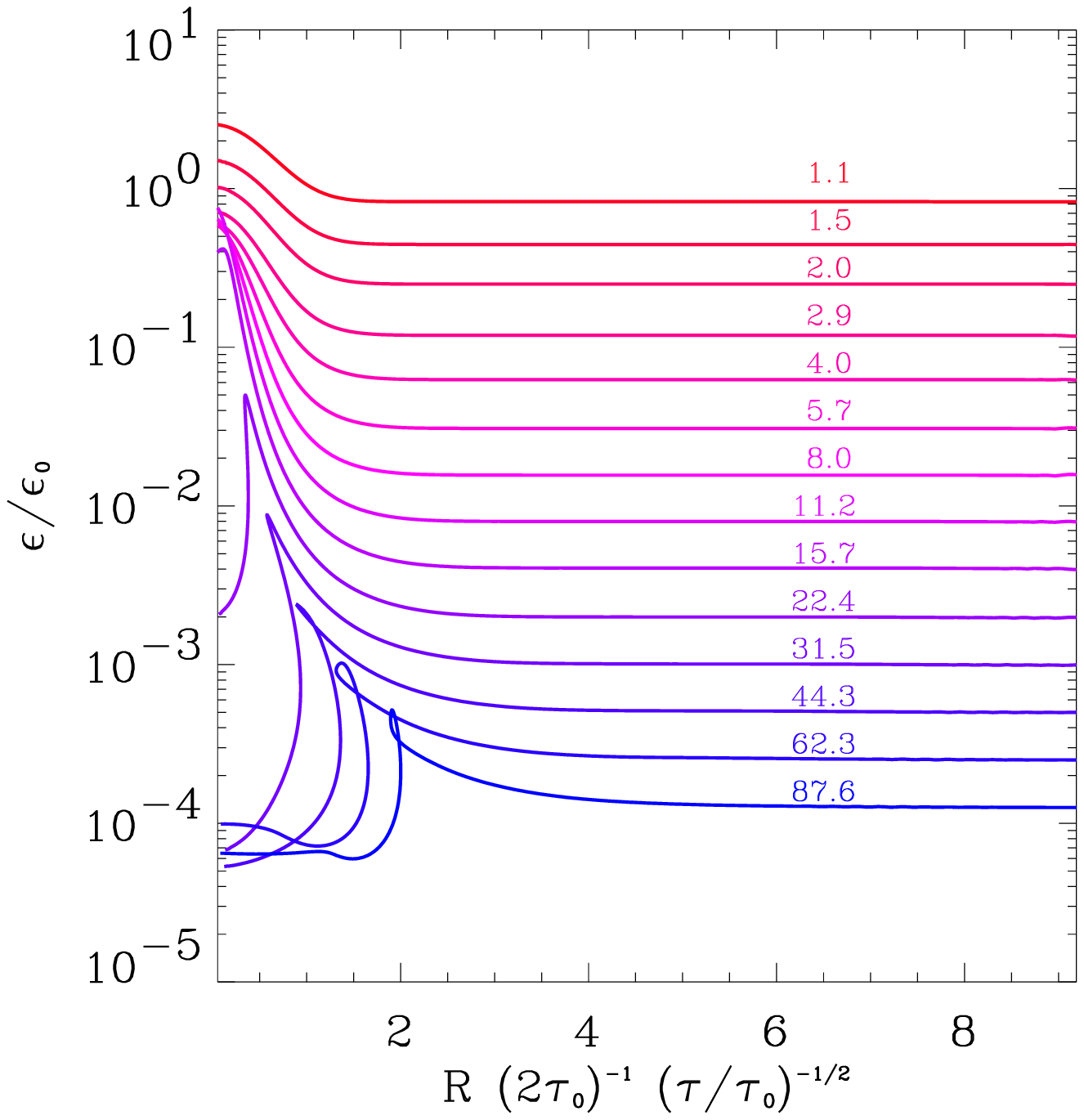}
\caption{\label{f4} Time evolution of an undercritical Gaussian-curve
perturbation with initial $\delta = 0.7006$. No black hole is formed.}
\end{minipage}
\end{figure}

Figures (\ref{f1}), (\ref{f2}), and (\ref{f3}) illustrate generic
features of the evolution of 
slightly supercritical perturbations for the three density
perturbation families, respectively. The curves
display the energy density, $\epsilon/\epsilon_0$, at constant proper
time, $\tau$, 
for each mass shell ($\tau$ is given in multiples of $t_0$ as labeled). 
In Hernandez--Misner coordinates, the lack of a well defined global
time variable corresponding to the cosmological FRW time at
infinity requires this local time slicing. As
described in \cite{baum95}, we integrate d$\tau = e^\Psi$d$u$ and store
$\tau(A,u)$ together with all other state variables. The curves are
then created by plotting the energy density along the isosurfaces of
$\tau$.
The radial coordinate is the circumferential radius, scaled such
that in the absence of a perturbation it may be associated with
cosmic comoving radius. Further, the initial horizon size, 
$\rh = 2 t_0$, is normalized to unity in the Figures.  
It is interesting to note that with this type of time slicing
$\epsilon(\tau = \mbox{const})$ may cease to be a single-valued 
function of the circumferential radius, $R$, in cases of strong curvature
(cf. Figure \ref{f4}). In particular, at the same proper time spheres
containing 
larger baryon number (labeled by $A$) may have smaller circumferential
radius than spheres containing smaller baryon number.

In all cases shown in Figures (\ref{f1}),(\ref{f2}), and (\ref{f3}), a black
hole with $\mbh 
\approx 0.37$ forms. The hydrodynamical evolution of the
three different perturbations exhibits strong similarities: initially,
the central overdensity grows in amplitude while
the outer underdensity, if present in the initial conditions, gradually
widens and levels out. A black hole forms in the interior. Some time
after the initial formation of an event horizon, material
close to the PBH but outside the event horizon bounces and launches a
compression wave traveling outward. This compression wave is connected
to the black hole by a rarefaction region that evacuates the
immediate vicinity of the black hole. The strength of the rarefaction
differs significantly for the Gaussian perturbation shape and the mass
compensated ones: while the latter display only a weak underdensity
that quickly equilibrates, the former gives rise to a drop in energy
density by three orders of magnitude. 

The bounce of material outside
the newly formed black hole is a feature intrinsic only to black holes
very close to the formation threshold. It effectively shuts off further
accretion of material onto the newly formed PBH. As Figure (\ref{f5})
demonstrates, no bounce occurs if the initial conditions are
sufficiently far above the threshold. Here, a large
black hole ($\mbh = 2.75$) forms whose event horizon reaches further out,
encompassing regions where the pressure gradient is smaller,
preventing pressure forces from overcoming gravitational
attraction.  Slightly below the PBH formation threshold, a
bounce occurs whose strength is proportional to the initial perturbation
amplitude (Figure \ref{f4}), indicating that the fluid bounce is
strongest for perturbations very close to the threshold. It is likely
that previous studies \cite{Nade} failed to
observe this phenomenon because their initial conditions were insufficiently
close to the black hole formation threshold. 
Their numerical simulations may also not have followed 
the hydrodynamical evolution for long enough after the initial 
formation of the PBH. 

\section{Accretion}
\label{accret}

Accretion onto PBHs and their resulting growth in mass has been a
highly debated subject since the suggestion that PBHs may grow in
proportion with the cosmological horizon mass \cite{PBHorigin1}. Both analytic 
\cite{analytic} and previous
numerical studies \cite{Nade} came to the conclusion that the growth of
PBH masses by ongoing accretion is negligible except, possibly,
for very contrived initial data for the perturbations. 

\begin{figure}[b]
\begin{minipage}{80mm}
\epsfxsize=80mm
\epsfbox{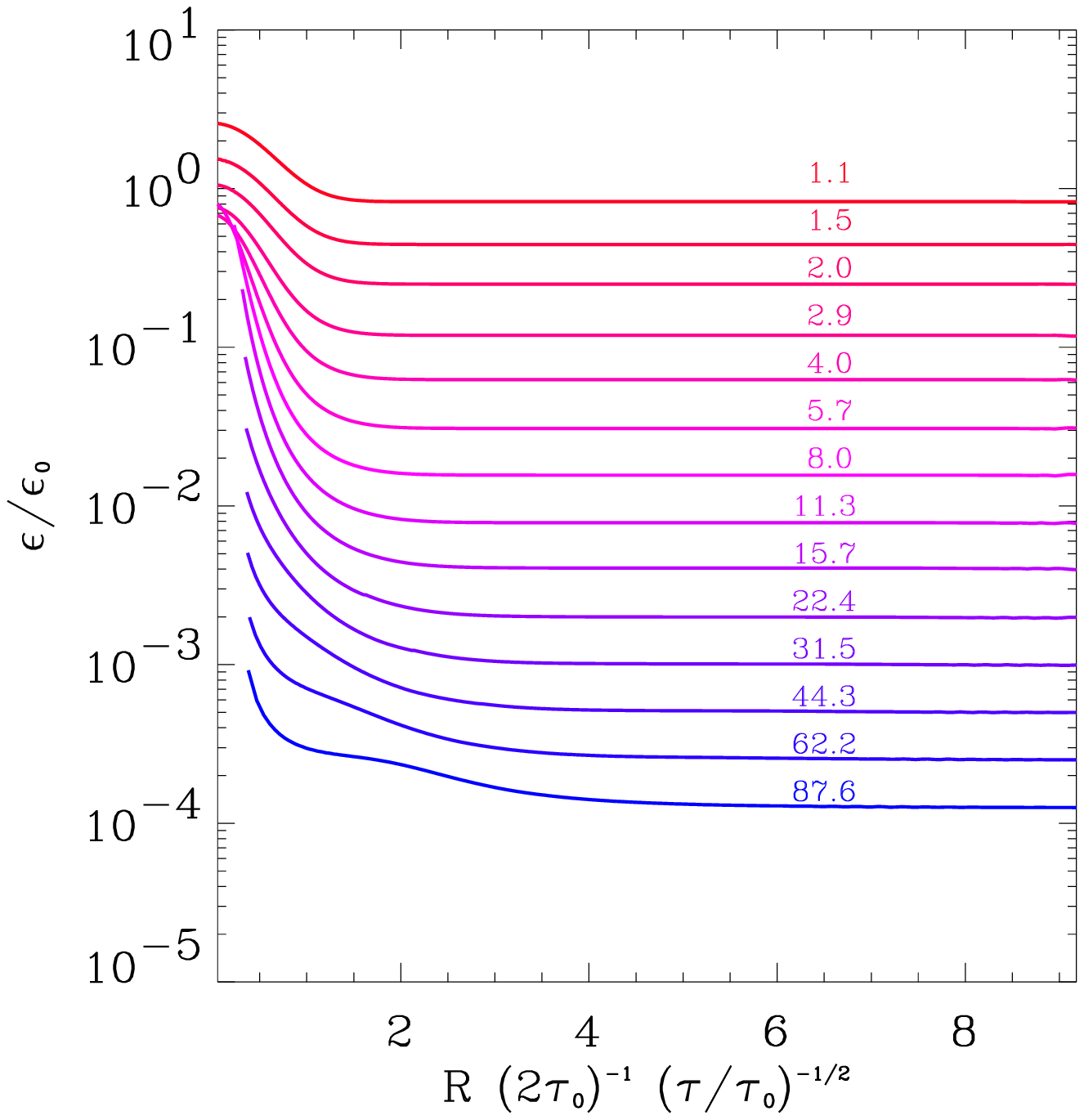}
\caption{\label{f5} Time evolution of an overcritical Gaussian-curve
perturbation with initial $\delta = 0.7196$, forming a 
black hole with mass $\mbh = 2.75$.}
\end{minipage}
\hspace{\fill}
\begin{minipage}{80mm}
\epsfxsize=80mm
\epsfysize=70mm
\epsfbox{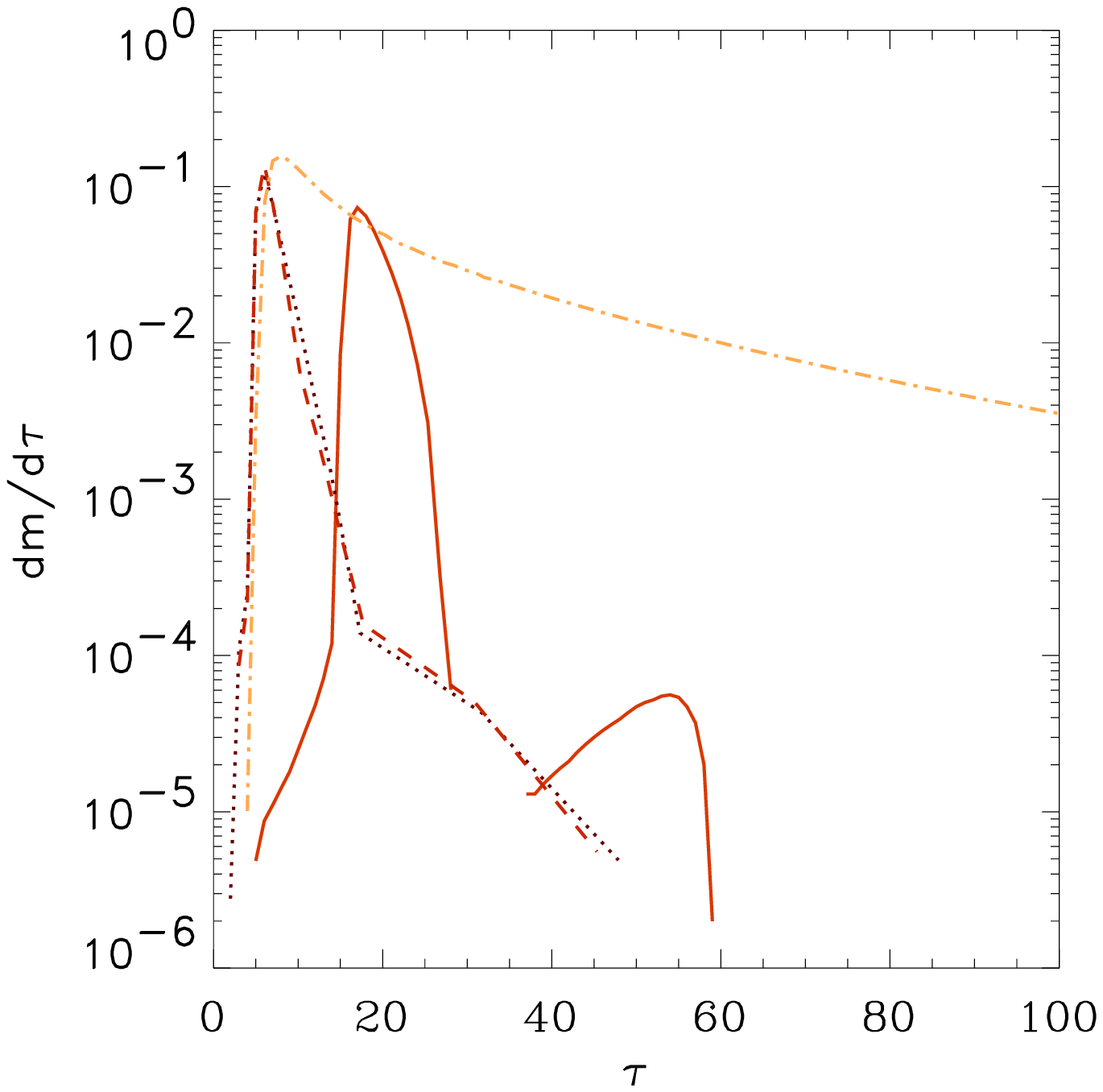}
\caption{\label{f6} Growth rate of $\mbh$ as a function of proper time
immediately 
outside of the event horizon, providing a measure for the accretion rate
onto the newly formed black hole. The lines correspond to the black
holes described in Figures \ref{f1} (dotted line), \ref{f2} (dashed
line), \ref{f3} (solid line), and \ref{f5} (dashed-dotted line).} 
\end{minipage}
\end{figure}
Our results generally
confirm this statement for small PBHs, but we find noticeable
differences between collapse simulations that exhibit the fluid
bounce and those that do not (Figure \ref{f6}). The rarefaction
following the outgoing density wave efficiently cuts off the flow of
material into the black hole. Comparing Figures (\ref{f6}) and
(\ref{f3}), it is recognized that the secondary phase of mass growth
for the Gaussian shape calculation may correspond to the rise of the
second wave crest
of the strongly damped density oscillation at the black hole event
horizon. This second rise in density is absent in the weaker bounces
of the Mexican-Hat and polynomial-shaped perturbation simulations. The 
large ($\mbh = 2.75$) black hole, on the other hand, continues to grow 
at a slowly decreasing rate for long times without gaining a
considerable amount of mass in the process. Based on these results, we
expect accretion to be insignificant for 
the determination of $\mbh$, at least for the types of perturbations
investigated here. 

\section{Scaling Relations for PBH Masses} 
\label{scaling}

Choptuik's discovery \cite{chop93} of critical phenomena in
gravitational collapse 
near the black hole formation threshold started an active and
fascinating line of research in numerical and analytical general
relativity (for recent reviews, see \cite{critrev}). For a
variety of matter models, it was found that the dynamics of
near-critical collapse
exhibits continuous or discrete self-similarity and power law scaling
of the black hole mass with the offset from the critical point
(Eq. \ref{scale}). In particular, Evans and Coleman \cite{evans94} found
self-similarity and mass scaling in numerical experiments of a
collapsing radiation fluid. They numerically determined the scaling exponent
$\gamma \approx 0.36$, followed by a linear perturbation analysis of
the critical solution by Koike \ea \cite{koike95} that yielded $\gamma \approx
0.3558$.

Until recently, it was believed that entering the scaling regime
requires a degree of fine-tuning of the initial data
that is unnatural for any astrophysical application. It was noted
\cite{niejed97} that fine-tuning to criticality
occurs naturally in the case of PBHs forming from a steeply declining
distribution of primordial density fluctuations, as generically predicted by
inflationary scenarios. In the radiation-dominated
cosmological epoch, the only difference with the fluid collapse
studied numerically by Evans and Coleman \cite{evans94} is the 
asymptotically expanding, finite
density background space--time of a FRW universe. Assuming that
self-similarity and mass scaling are consequences of an intermediate
asymptotic solution that is independent of the asymptotic boundary
conditions, Eq. (\ref{scale}) is applicable to PBH masses,
allowing the derivation of a universal PBH initial mass function
\cite{niejed97}. Furthermore, cosmological constraints based on evaporating
PBHs are slightly modified as a consequence of 
the production of not only horizon-size PBHs, as previously assumed,
but the additional production of smaller,
sub-horizon mass black holes at each epoch \cite{yok98}.   

\begin{figure}
\epsfxsize=80mm
\epsfbox{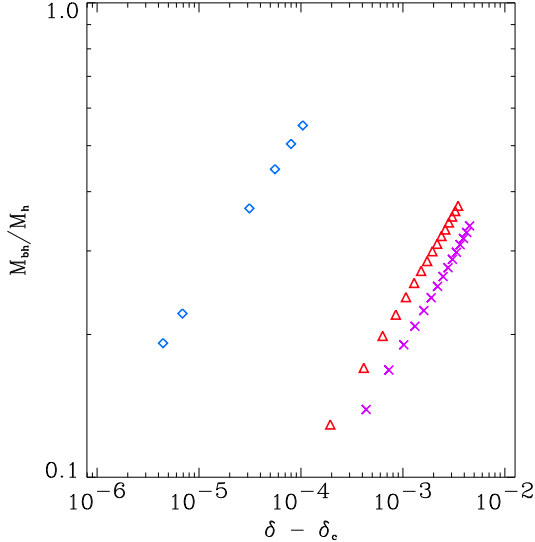}
\caption{\label{f7} Black hole masses as a function of the distance to
the formation threshold, $\delta - \dc$, for three different
perturbation shape families. Using the smallest six black holes of
each family, the best fit parameters to Eq. (\ref{scale}) are:
$\gamma = 0.36$, $K = 2.85$, $\dc = 0.6745$ (Mexican-Hat
perturbation, triangles); $\gamma = 0.37$, $K = 2.39$, $\dc = 0.7122$
(polynomial perturbation, crosses); $\gamma = 0.34$, $K = 11.9$,
$\dc = 0.7015$ (Gaussian-curve perturbation, diamonds).}
\end{figure}
Figure (\ref{f7}) presents numerical evidence that mass scaling
according to Eq. (\ref{scale}) occurs in the collapse of
near-critical black holes in an asymptotic FRW space--time, and
therefore applies to PBH formation. All three perturbation families give
rise to scaling solutions with a scaling exponent $\gamma
\approx 0.36$. Only the smallest six black holes of all families were
included to obtain the numerical best fit quoted in the figure
captions. On larger mass scales, deviations from mass scaling with a
fixed exponent become noticeable; in all cases, $\gamma$ tends to
increase slightly for larger $\mbh$. Owing to resolution limitations
discussed in Section (\ref{numerics}), we were unable to compute the
formation of smaller black holes than the ones shown in Figure
(\ref{f7}). 

To linear order, the scaling relation (\ref{scale}) is invariant under
transformations of the control parameter $\delta$ up to a change of
the coefficient $K$. This was tested explicitly by choosing different
definitions of $\delta$ (the perturbation amplitude $A$, the total
excess mass for the Gaussian-shaped perturbation, 
and the excess mass within the horizon volume) for the
numerical fit and obtaining identical values for $\gamma$.

\section{Conclusions}
In the general framework of primordial black hole (PBH) formation
from horizon-size, pre-existing density perturbations, we
numerically solved the spherically symmetric general 
relativistic hydrodynamical equations in order to
study the collapse of radiation fluid overdensities in an
expanding Friedmann--Robertson--Walker (FRW) universe. The algorithm is
adopted from an implementation of the Hernandez--Misner coordinates
\cite{hermis66} by Baumgarte \ea \cite{baum95}. It allows the
convenient computation of 
black hole formation and superhorizon scale dynamics by virtue of its
time coordinate, chosen to be constant along outgoing null surfaces. 

One of the parameters entering the statistical analysis of
cosmological consequences and constraints due to the possible
abundant production of PBHs is the threshold parameter, $\dc$, 
corresponding to the amplitude of the smallest perturbations that still
collapse to a black hole. It generally depends on the specific
perturbation shape at the time of horizon crossing. We studied three
generic families of energy density perturbations, one with a finite
total excess mass
with respect to the unperturbed FRW solution and two mass compensated
ones. Defining the control parameter, $\delta$, as the total excess
gravitational mass of the perturbed space--time with respect to the
unperturbed FRW background enclosed in the initial horizon volume, our
calculations yield a similar threshold value for all three fluctuation
shape families, $\dc \approx 0.7$. 

We investigated features of collapsing space--times very close
to the threshold of black hole formation embedded in an expanding FRW
solution. If the initial perturbation
is smaller than $\dc$, it grows until pressure forces at the origin
cause the fluid to bounce, creating an outgoing pressure wave
followed by a rarefaction, but no black hole. Initial conditions
slightly exceeding the threshold, on the other hand, lead to the
formation of a very small black hole at the origin; however, the
pressure gradient immediately outside of the event horizon is still
sufficiently steep to force the fluid to bounce. The launch of a
compression wave can be observed in simulations of all three perturbation
shapes. It is strongest in the case of a pure initial overdensity,
parameterized here as a Gaussian curve, where the density behind the
pressure wave drops by three orders of magnitude. Increasing $\delta$
to values significantly above $\dc$, the bounce becomes weaker and
finally disappears, signaling the failure of the pressure gradient at
the event horizon to overcome gravitational attraction.

This behavior has important consequences for the accretion onto
PBHs immediately after their formation. If a bounce occurs, the inner
rarefaction shuts off accretion almost completely before any significant amount
of material has been accreted. On the other hand, 
black holes that form from sufficiently
large overdensities, where a bounce is suppressed, may accrete 
at a slowly decreasing rate for
a long time. Since most PBHs created from collapsing primordial density
fluctuations with a steeply declining amplitude distribution form very
close to $\dc$ \cite{niejed97}, we conclude that accretion is unimportant
for the estimation of PBH masses. This is in agreement with previous
studies \cite{Nade}, albeit for different reasons. 

Finally, the previously suggested \cite{niejed97} scaling relation between
$\mbh$ and $\delta - \dc$, based on the analogy with critical
phenomena observed in near-critical black hole collapse in
asymptotically non-expanding space--times \cite{chop93}, was confirmed
numerically for an asymptotic FRW background. For the smallest black
holes in our investigation, the scaling exponent is $\gamma \approx
0.36$, which is identical to the non-expanding numerical and
analytical result \cite{evans94} within our numerical accuracy. The
parameter $K$ of Eq. (\ref{scale}), needed to evaluate the PBH
initial mass function derived in \cite{niejed97}, 
was found to range from $K \approx 2.4$ to $K \approx 12$.  

We wish to thank T. Baumgarte for providing the original version of
the hydrodynamical code, and T. Abel, A. Olinto, and V. Katalini\'{c}
for helpful discussions. Part of this research was supported by an
Enrico-Fermi-Fellowship.


\begin{thebibliography}{99}
\bibitem{com1} Furthermore, we use the same notation and geometrized
units as \cite{baum95}.
\bibitem{com2}
We define PBH mass as a function of time 
as the rest mass-energy interior to the event horizon at the proper time of 
an infalling fluid element which just passes over the event horizon.
Since the lapse of proper time between the fluid
element first entering the strong gravitational field of the PBH and 
the fluid element crossing over the event horizon is generally short,
this definition is stable in terms of our specific 
identification of the event horizon. Given as a function of time of a
distant observer, on the other hand, 
strong gravitational time dilation effects render the PBH mass
extremely sensitive to the definition of the event horizon in terms of
the lapse function.
\end{thebibliography}
\end{document}